\documentclass[prd,aps,reprint,noeprint,superscriptaddress,amsmath, amssymb]{revtex4-1}
\usepackage{graphicx}% Include figure files
\usepackage{subfigure}
\usepackage{lineno}
\usepackage{epsfig}
\usepackage{multirow}
\usepackage{overpic}
\usepackage{color}
\usepackage{bm} %用于加粗公式中的符号
\usepackage{xspace} %空格相关的
\usepackage{dcolumn}% Align table columns on decimal point
\usepackage{booktabs,tabularx}
\usepackage{array}
\usepackage{textcomp}
\usepackage[colorlinks,linkcolor=blue,urlcolor=blue,citecolor=blue]{hyperref}
\usepackage{tikz}

\usepackage{cleveref}%multiple figure reference

\newcommand{\BESIIIorcid}[1]{\href{https://orcid.org/#1}{\hspace*{0.1em}\raisebox{-0.45ex}{\includegraphics[width=1em]{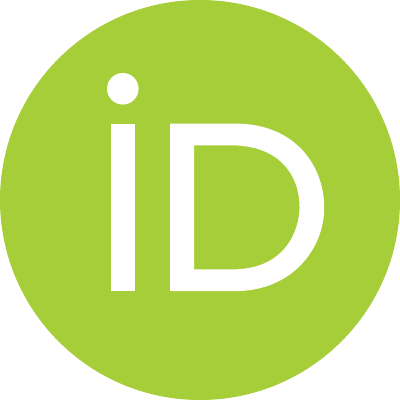}}}}
\begin{document}
\normalsize
\parskip=5pt plus 1pt minus 1pt

%%%%%%%%%%%%%%%%%%%%%%%%%%%%%%%%%%%%%%%%%%%%%%%%%%%%%%%%%%%%%%%%%

\title{
    \boldmath  Observation of $\bar{\Lambda}p\to K^{+}\pi^{+}\pi^{-}\pi^{0}$ and $\bar{\Lambda}p\to K^{+}\pi^{+}\pi^{-}2\pi^{0}$
    % \boldmath  Observation of $\bar{\Lambda}p\to K^{+}\pi^{+}\pi^{-}+k\,\pi^{0}$ ($k=1,2$)%with $\bar{\Lambda}$ from $J/\psi$ Decays at BESIII
}

\author{%% Saved at => 2025-10-14
M.~Ablikim$^{1}$\BESIIIorcid{0000-0002-3935-619X},
M.~N.~Achasov$^{4,d}$\BESIIIorcid{0000-0002-9400-8622},
P.~Adlarson$^{81}$\BESIIIorcid{0000-0001-6280-3851},
X.~C.~Ai$^{87}$\BESIIIorcid{0000-0003-3856-2415},
C.~S.~Akondi$^{31A,31B}$\BESIIIorcid{0000-0001-6303-5217},
R.~Aliberti$^{39}$\BESIIIorcid{0000-0003-3500-4012},
A.~Amoroso$^{80A,80C}$\BESIIIorcid{0000-0002-3095-8610},
Q.~An$^{77,64,\dagger}$,
Y.~H.~An$^{87}$\BESIIIorcid{0009-0008-3419-0849},
Y.~Bai$^{62}$\BESIIIorcid{0000-0001-6593-5665}
% ...
\\
\vspace{0.2cm}
(BESIII Collaboration)\\
\vspace{0.2cm} {\it
$^{1}$ Institute of High Energy Physics, Beijing 100049, People's Republic of China\\
% ...
$^{4}$ Budker Institute of Nuclear Physics SB RAS (BINP), Novosibirsk 630090, Russia\\
% ...
$^{31}$ INFN Sezione di Ferrara, (A)INFN Sezione di Ferrara, I-44122, Ferrara, Italy; (B)University of Ferrara, I-44122, Ferrara, Italy\\
% ...
$^{39}$ Johannes Gutenberg University of Mainz, Johann-Joachim-Becher-Weg 45, D-55099 Mainz, Germany\\
% ...
$^{62}$ Southeast University, Nanjing 211100, People's Republic of China\\
% ...
$^{64}$ State Key Laboratory of Particle Detection and Electronics, Beijing 100049, Hefei 230026, People's Republic of China\\
% ...
$^{77}$ University of Science and Technology of China, Hefei 230026, People's Republic of China\\
% ...
$^{80}$ University of Turin and INFN, (A)University of Turin, I-10125, Turin, Italy; (B)University of Eastern Piedmont, I-15121, Alessandria, Italy; (C)INFN, I-10125, Turin, Italy\\
$^{81}$ Uppsala University, Box 516, SE-75120 Uppsala, Sweden\\
% ...
$^{87}$ Zhengzhou University, Zhengzhou 450001, People's Republic of China\\
\vspace{0.2cm}
$^{\dagger}$ Deceased\\
% ...
$^{d}$ Also at the Novosibirsk State University, Novosibirsk, 630090, Russia\\
% ...
}
%% ends here %%
}
%%%%%%%%%%%%%%%%%%%%%%%%%%%%%%%%%%%%%%%%%%%%%%%%%%%%%%%%%%%%%%%%%
\begin{abstract}
    Using $(10087 \pm 44) \times 10^6$ $J/\psi$ events collected with the BESIII detector at a center-of-mass energy of $\sqrt{s}=3.097$ GeV,
    the antihyperon-nucleon annihilation processes $\bar{\Lambda} p \to K^+ \pi^+ \pi^- + k\pi^0$ ($k=1,2,3$)
    are studied at an incident $\bar{\Lambda}$ momentum of approximately 1.074~GeV/$c$.
    The reactions $\bar{\Lambda} p \to K^+ \pi^+ \pi^- \pi^0$ and $\bar{\Lambda} p \to K^+ \pi^+ \pi^- 2\pi^0$ are observed for the first time,
    with corresponding cross sections $\sigma_{\bar{\Lambda} p \to K^+ \pi^+ \pi^- \pi^0} = 8.5^{+1.2}_{-1.1}~(\rm{stat.}) \pm 0.4~ (\rm {syst.})$~mb
    and $\sigma_{\bar{\Lambda} p \to K^+ \pi^+ \pi^- 2\pi^0} = 7.9^{+1.9}_{-1.7} \pm 0.4$~mb.
    No significant signal is found for $\bar{\Lambda} p \to K^+ \pi^+ \pi^- 3\pi^0$, and an upper limit of 7.2 mb is set at a 90\% confidence level.
    An evidence for the $K^{*}(892)^+$ resonance is seen in the $K^+\pi^0$ invariant mass spectrum $M_{K^+\pi^0}$ for $k=1$,
    and the corresponding cross section for $\bar{\Lambda} p \to K^{*}(892)^+ \pi^+ \pi^-$ is measured to be
    $\sigma_{\bar{\Lambda} p \to K^{*}(892)^+ \pi^+ \pi^-} = 12.5^{+3.8}_{-3.4} \pm 1.2$ mb.
    Owing to the limited statistics, possible interference effects are not considered.
    These findings offer crucial input to deepen our understanding of the antihyperon-nucleon interactions.
\end{abstract}

\maketitle
%\linenumbers

One goal of low-energy Quantum Chromodynamics (QCD) is to provide a unified and quantitative description of hadron-hadron interactions.
For the nucleon-nucleon system, decades of precise scattering data have enabled high-accuracy potential models and chiral effective field theory (chiral EFT) descriptions that underpin modern nuclear structure and reaction theory,
with growing support from lattice QCD~\cite{Machleidt2011,Epelbaum2009,Wiringa1995}.
Extending this understanding to systems with strangeness is essential for testing SU(3)-flavor aspects of the strong interaction,
constraining hypernuclear structure and spectroscopy, and determining the equation of state (EoS) of dense baryonic matter.
The latter is believed to play an important role in the so-called hyperon puzzle of neutron stars~\cite{Oertel2017,Demorest2010,Antoniadis2013,Cromartie2020,Lonardoni2015}.
Considerable progress has been made in the hyperon-nucleon (YN) sector through chiral EFT, SU(3)-based potentials, and lattice QCD,
together with input from scattering, hypernuclear spectroscopy,
and femtoscopy measurements~\cite{Haidenbauer2013,Gal2016,Rijken2010,Tolos2020}.
In contrast, the antihyperon-nucleon ($\rm \bar{Y}N$) interaction remains poorly known,
with only limited theoretical and experimental constraints available to date, though related insights can be indirectly gained from studies of $p\bar{p}$ annihilation to $\rm \bar{Y}Y$~\cite{Barnes:1987aw, Barnes:1990wh, Timmermans:1992fu, Haidenbauer:1992hv, Kohno:1987uj}. 

Extensive studies of nucleon-antinucleon interactions, from low-energy $\bar{p}p$ and $\bar{p}\rm N$
scattering to antiprotonic atoms, show a strongly absorptive,
short-range interaction well described by complex optical potentials and characterized by prolific multi-meson annihilation channels~\cite{Friedman2007}.
Theoretical investigations also emphasize that experimental measurements of the $\rm \bar{Y}N$ annihilation cross-sections provide unique insights that are complementary to those obtained from YN interactions. While YN scattering primarily probes the real part of the optical potential, the $\rm \bar{Y}N$ annihilation channels offer a direct constraint on the imaginary part, representing the absorption mechanism in the nuclear medium~\cite{Friedman2007}. These data are crucial for testing the validity of $G$-parity transformation in the strangeness sector.  %and for distinguishing between ``deep" and ``shallow" potential models. 
Furthermore, the annihilation dynamics at low energies serve as a sensitive probe for the partial restoration of chiral symmetry and many-body effects in dense hadronic matter, which are essential for refining theoretical models of hyperon-nuclear systems~\cite{Friedman2007}.
Although these systems provide benchmarks for antibaryon-matter interactions,
direct constraints on $\rm \bar{Y}N$ dynamics are hindered by the lack of effective antihyperon beams.
By analogy, near threshold one expects similarly strong absorption and multi-meson annihilation in the strange sector, rendering the effective $\rm \bar{Y}N$ interaction highly complex in nuclear matter. Such dynamics have implications for transport and hadronization in heavy-ion collisions,
strange-antibaryon survival, light antinuclei formation and breakup, and the use of antibaryons as probes of nuclear media~\cite{Tolos2020,Friedman2007}.
To date, elastic scattering $\bar{\Lambda} p\to \bar{\Lambda} p$ is the only measured $\rm \bar{Y}N$ process~\cite{BESIII:2024geh};
no inelastic or annihilation data exist.
Therefore, dedicated $\rm \bar{Y}N$ scattering measurements are urgently needed to deepen our understanding of the $\rm \bar{Y}N$ interaction and provide
complementary insights on the YN interaction.

\begin{figure}[htbp]
    \begin{center}
    \begin{overpic}[width=0.5\textwidth]{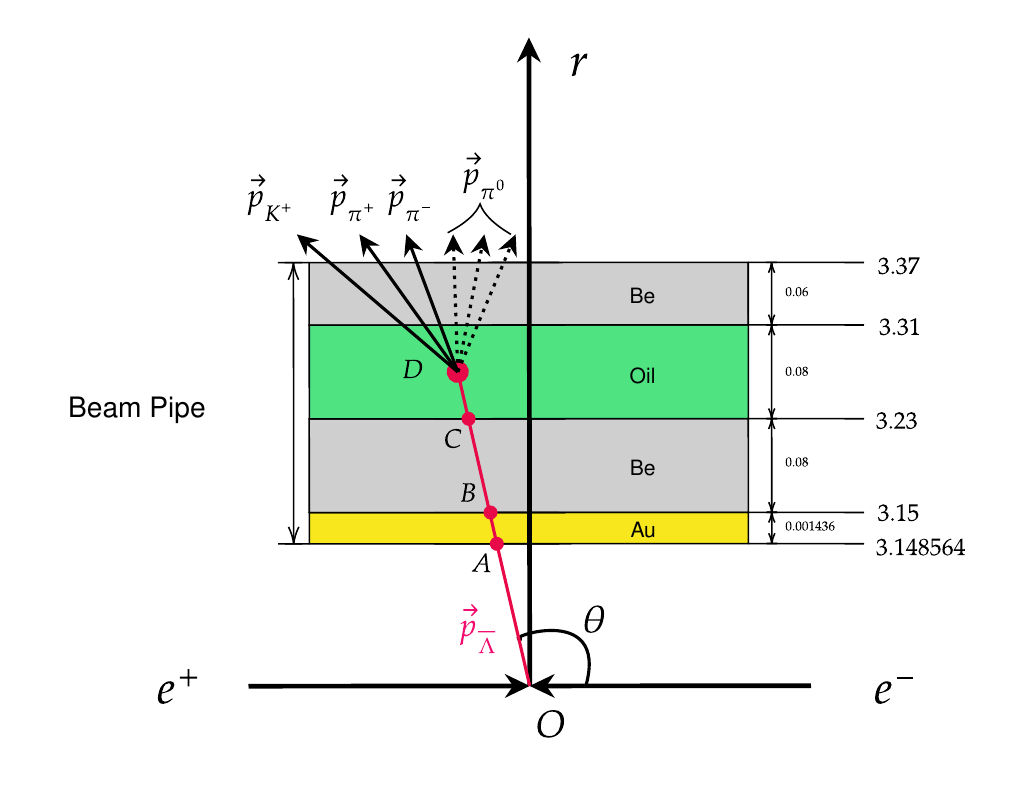}
    \end{overpic}
    \end{center}
    \vspace{-0.9cm}
    \caption{
        Schematic diagram of the beam pipe and the $\bar{\Lambda}$ annihilation trajectory inside the target; the length units are centimeters (cm).
        $O$ is the interaction point of the $e^+ e^-$ collision.
        The horizontal axis is the $e^+e^-$ beam line, and the vertical axis ($r$ axis) denotes the transverse distance from the beam line.
        $\theta$ is the angle between the incident $\bar{\Lambda}$ antihyperon and the $z$ axis.
    }
    \label{fig:beampipe}
\end{figure}

Electron-positron facilities offer a unique path forward by producing hyperon-antihyperon pairs with well-defined kinematics in
two-body decays.
At BESIII, $J/\psi \to \Lambda\bar{\Lambda}$ decays deliver an intense,
quasi-monoenergetic flux of $\bar{\Lambda}$ hyperons with momentum $1.074~\mathrm{GeV}/c$ and
a spread of $\pm 0.017~\mathrm{GeV}/c$, arising from the $\pm 11~\mathrm{mrad}$ horizontal crossing angle of the $e^{+}e^{-}$ beams,
providing a clean and intensive source of antihyperons~\cite{Li2017haibo}.
In this Letter, we present the study of the
$\bar{\Lambda} p \to K^{+}\pi^{+}\pi^{-} + k\,\pi^{0}$ reactions with $k=1,2,3$ and search for possible intermediate resonances in the final states,
using a recently developed technique~\cite{Dai:2024myk,Yuan:2021yks}, which has been employed in recent pioneering BESIII analyses~\cite{BESIII:2023clq,BESIII:2023trh,BESIII:2024geh,BESIII:2025bft}.
To study the reactions of interest, a data set of $(10087 \pm 44)\times 10^{6}$ $J/\psi$ events~\cite{BESIII:2021cxx} recorded with the BESIII detector
is used. % to study $\bar{\Lambda}$ antihyperons from $J/\psi \to \Lambda\bar{\Lambda}$ decays.
Details of the BESIII detector, a magnetic spectrometer operating at the Beijing Electron-Positron Collider (BEPCII)~\cite{Yu:2016cof},
are given in Refs.~\cite{Ablikim2010,BESIII:2020nme,Li:2017eToF,Guo:2017eToF,Cao:2020ibk}.
As the $\bar{\Lambda}$ antihyperons traverse the beam pipe,
they can interact with the surrounding material, which is composed of $^{197}\mathrm{Au}$, $^{9}\mathrm{Be}$,
and a cooling oil with composition $^{12}\mathrm{C}:{}^{1}\mathrm{H}=1\!:\!2.13$ (see Fig.~\ref{fig:beampipe}),
producing multi-meson final states with strangeness S = 1 that directly encode short-range $\rm \bar{Y}N$ dynamics and annihilation topologies.
In particular, by using the at-rest protons ($^{1}\mathrm{H}$) in the cooling oil as targets, $\bar{\Lambda}p$ annihilation
can be directly probed. This ensures a well-defined initial state for signal identification, whereas annihilation on bound nucleons in heavier nuclei would be smeared by Fermi motion.

Using a {\sc geant4}-based~\cite{G42002iii} Monte Carlo (MC) package,
simulated samples are produced, incorporating the geometric description of the BESIII detector,
the propagation of particles through the detector and the detector response.
An inclusive MC sample containing 10 billion generic $J/\psi$ decays is used to investigate potential backgrounds.
The formation of the $J/\psi$ resonance is simulated by the MC event generator {\sc kkmc}~\cite{Jadach2001},
where the beam-energy spread and initial-state radiation in the $e^+ e^-$ annihilation are taken into account.
The known decay modes are generated by {\sc evtgen}~\cite{Lange2001,*Ping2008} using branching fractions taken from
the Particle Data Group (PDG)~\cite{PDG}, while the unknown decay modes are modeled with {\sc lundcharm}~\cite{Chen2000,*Yang2014a}.
The signal processes considered in this analysis are $J/\psi \to \Lambda \bar{\Lambda}$ with $\bar{\Lambda} p \to K^{+}\pi^{+}\pi^{-} + k\,\pi^{0}$, with $k=1,2,3$
and $\bar{\Lambda} p \to K^{*}(892)^{+}\pi^{+}\pi^{-}$, $K^{*}(892)^{+}\to K^+ \pi^0$, $\pi^0\to \gamma \gamma$.
In the signal simulation, the angular distribution of $J/\psi\to\Lambda\bar{\Lambda}$ is generated according to the measurement in Ref.~\cite{BESIII:2022qax}, and the mass and the width of $K^{*}(892)^{+}$ are set to be $891.67\pm0.26$~MeV/$c^2$ and $51.4\pm0.8$~MeV/$c^2$~\cite{PDG}.
The reactions $\bar{\Lambda}p\to  K^{+}\pi^{+}\pi^{-} + k\,\pi^{0}$ are simulated with the proton at rest,
with the angular distributions of the final state mesons following an isotropic phase-space distribution. The efficiency differences among the various data-taking periods are taken into account in the MC simulation, and the same event selection criteria are applied to both data and MC samples in this Letter.

To investigate the signal process, we tag the $\Lambda$ hyperon via its decay mode $\Lambda \to p \pi^-$ (single-tag (ST)),
from which we determine the four-momentum of the $\bar{\Lambda}$ antihyperon by missing kinematics.
The signal processes are then reconstructed from the remaining tracks and neutral showers in the BESIII calorimeter not used in the tag selection.
Candidate events in which the $\bar{\Lambda}$ annihilates with a proton into $K^{+}\pi^{+}\pi^{-} + k\,\pi^{0}$
and the $\Lambda$ decays into the tag mode are called secondary annihilation (SA) events.
The cross section of the signal process is given by
\begin{equation}
\label{equ:crosssection}
    \sigma = \frac{M}{N_{A} \cdot \rho_{T} \cdot l} \cdot \frac{N_{\rm SA}}{\epsilon_{\rm sig} \cdot N_{\rm ST}} \cdot \frac{1}{{\mathcal B}^{k}_{\pi^0 \to \gamma\gamma}},
\end{equation}
where $M=1.00794$~g/mol and $\rho_T = 0.1221$~g$\cdot$cm$^{-3}$ are the molar mass and density of hydrogen in the cooling oil, respectively,
and ${N}_{\rm ST}$ and ${N}_{\rm SA}$ are the ST and SA yields.
Here, $\epsilon_{\rm sig} = \epsilon_{\rm SA}/\epsilon_{\rm ST}$ denotes the signal efficiency in the presence of
an ST $\Lambda$ hyperon, with $\epsilon_{\rm ST}$ and $\epsilon_{\rm SA}$ denoting ST and SA efficiencies, respectively,
$N_A = 6.02214076 \times 10^{23}$~mol$^{-1}$ is Avogadro’s number~\cite{Tiesinga:2021myr},
and the parameter $l = 0.5509$~mm is the average path length of the incident $\bar{\Lambda}$ in the beam pipe cooling oil.
The latter is denoted ``CD" in Fig.~\ref{fig:beampipe},
and can be extracted from MC simulations with high precision.
$\mathcal{B}^k_{\pi^0 \to \gamma \gamma}$ is the branching fraction of the decay $\pi^0 \to \gamma \gamma$, raised to the $k$-th power,
where $\mathcal{B}_{\pi^0 \to \gamma \gamma} = (98.823\pm0.034)\%$~\cite{PDG}.

Charged tracks detected in the multilayer drift chamber (MDC) are required to be within a polar angle
range of $|\rm{cos\theta}|<0.93$, where the polar angle $\theta$ is defined with respect to the $z$-axis,
which is the symmetry axis of the MDC.
According to the MC simulation, the momenta of the proton and pion from $\Lambda$ decay are well separated: 
all pions have momenta below 500 MeV/c, while protons have momenta above 500 MeV/c.
Thus, on the tag side, the charged tracks with momentum larger than 500 MeV/c are considered as proton candidates, and
otherwise as pion candidates. 

The $\pi^0$ candidates are formed by $\gamma\gamma$ pairs, where the photons are identified from showers in the electromagnetic calorimeter (EMC).
The deposited energy of each shower is required to be larger than 25~MeV in the barrel region ($|\cos\theta| < 0.80$) and larger than 50~MeV in the end-cap region ($0.86 < |\cos\theta| < 0.92$), following the standard BESIII photon selection criteria~\cite{BESIII:2023edk, BESIII:2021rqk}.
To suppress electronic noise and energy deposits unrelated to the event, the EMC time of the photon candidate relative to the event start time is required to be within [0, 700]~ns, consistent with the BESIII standard selection~\cite{BESIII:2023edk, BESIII:2021rqk}.
The $\gamma\gamma$ pairs with invariant mass in the range $[0.115, 0.150]$~GeV/$c^2$ are regarded as $\pi^0$ candidates.
A kinematic fit constraining the $\gamma\gamma$ invariant mass to the nominal $\pi^0$ mass~\cite{PDG} (1C fit) is performed, and the updated four-momentum is used in further analysis.

\begin{figure}[htbp]
    \begin{center}
        \begin{overpic}[width = 0.8\linewidth]{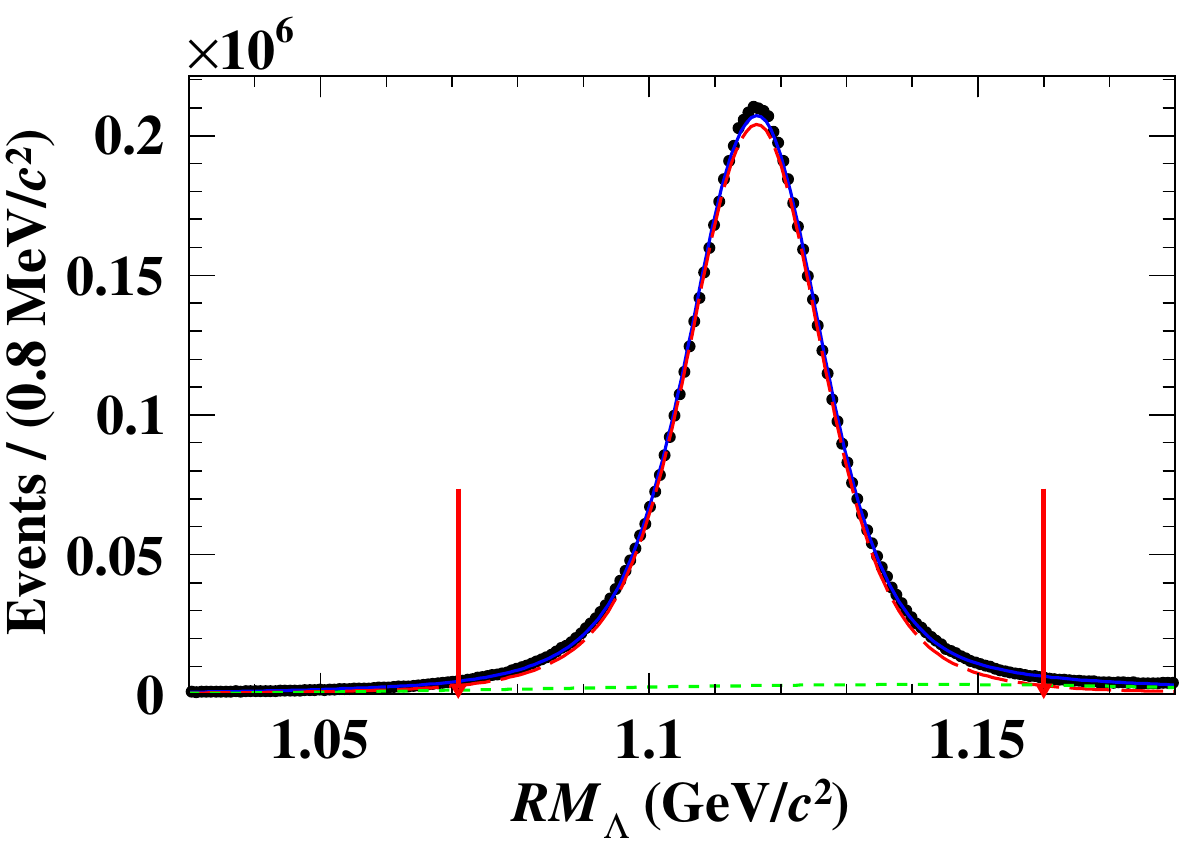}
        \end{overpic}
    \end{center}
    \vspace{-0.6cm}
    \caption{
        Fit to the $RM_{\Lambda}$ distribution for the ST sample.
        The dots with error bars are data, the blue solid line is the total fit,
        the red long-dashed and green short-dashed lines denote the fitted signal and background shapes.
    }
    \label{fig:SingleTag}
\end{figure}

To reconstruct the tagged $\Lambda$ candidates, a vertex fit is performed using all combinations of $p$ and $\pi^-$ candidates, constraining them to a common vertex. The combinations are required to satisfy $\chi^2 < 200$ for the vertex fit, where the number of degrees of freedom is 1. In addition, the invariant mass of the $p\pi^-$ system is required to be within $[1.110, 1.121]$~GeV/$c^2$.
If there are multiple $\Lambda$ candidates, only the $p \pi^-$ combination with the minimum vertex-fit $\chi^2$ is kept.
The ST yield is then extracted from an unbinned maximum-likelihood fit
on the recoil-mass spectrum against the reconstructed $\Lambda$ candidate ($RM_{\Lambda}$),
which is defined as
\begin{equation}
    RM_{\Lambda} =
    \sqrt{(E_{\rm cms}-E_{p \pi^-})^2/c^4 - \vec{p}_{p\pi^-}^2/c^2}.
    \label{Eq:mRec}
\end{equation}
Here, $E_{\rm cms}$ is the center-of-mass energy, $E_{p \pi^-}$ and $\vec{p}_{p \pi^-}$
are the energy and momentum of the tagged $p \pi^-$ in the $J/\psi$ rest frame, respectively.
In the fit, the signal is modeled using the shape of  MC samples convolved with a Gaussian function,
which accounts for the difference
in the mass resolution between data and MC simulation.
The background shape is described by a third-order Chebyshev polynomial.
The fit to the $RM_{\Lambda}$ distribution is shown in Fig.~\ref{fig:SingleTag};
the red arrows denote the signal region, $RM_{\Lambda}$ $\in$ [1.071, 1.160]~GeV/$c^2$,
and the corresponding ST yield is determined to be $N_{\rm ST} = (7511.4 \pm 3.6) \times 10^{3}$.

\begin{figure*}[htbp]
    \begin{center}
        \includegraphics[width=0.32\textwidth]{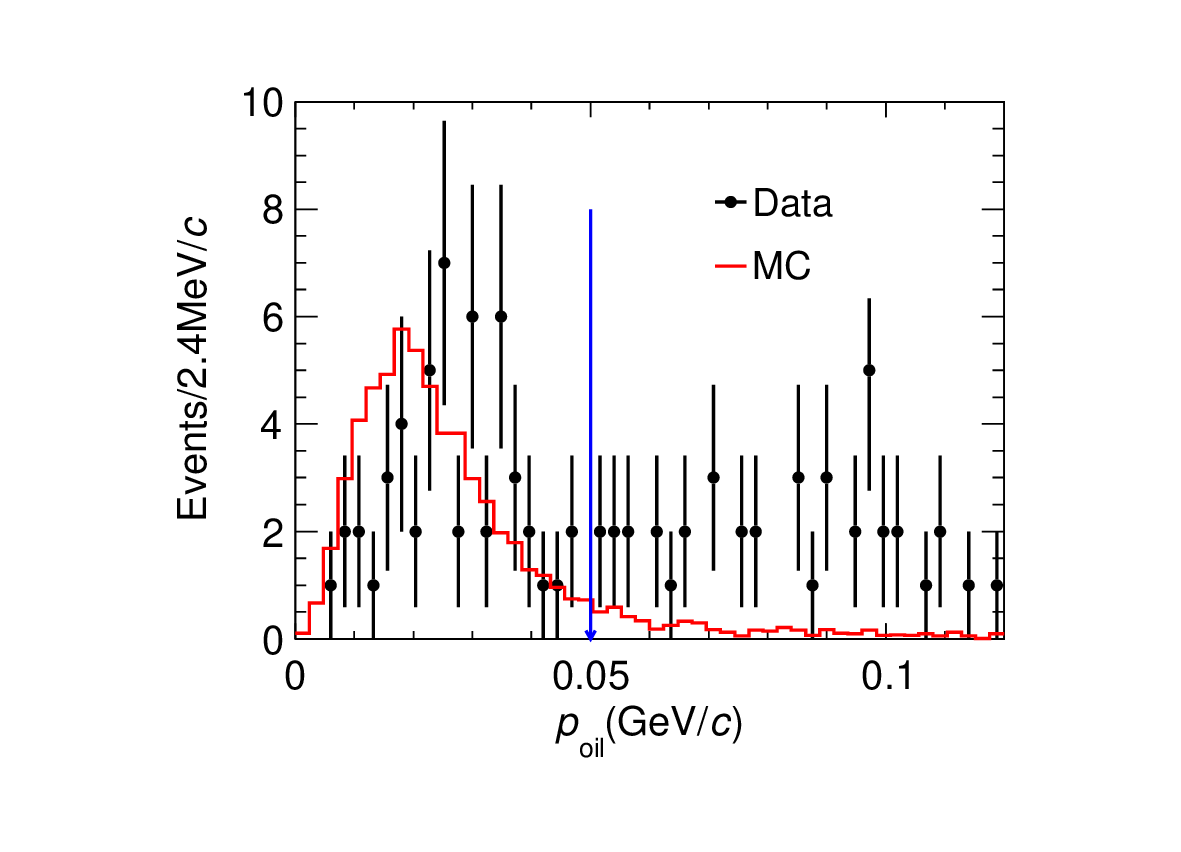}
        \includegraphics[width=0.32\textwidth]{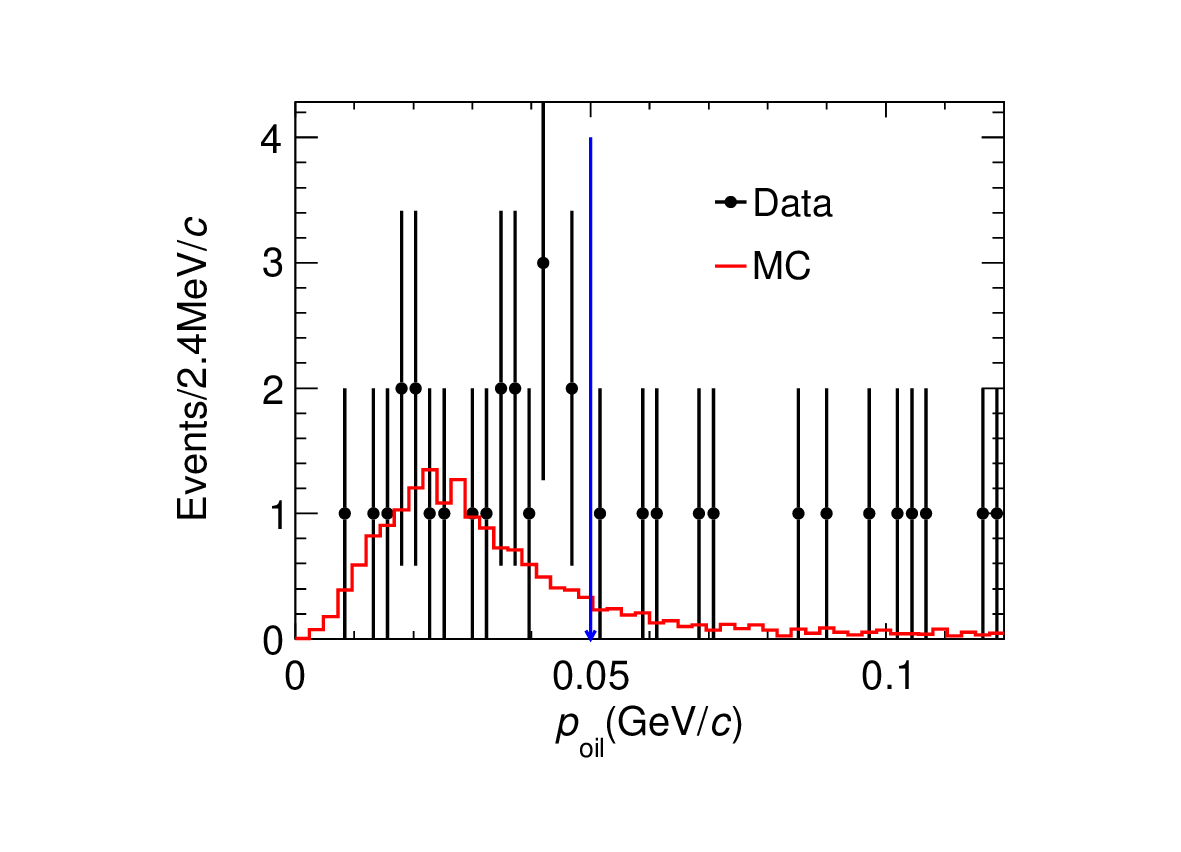}
        \includegraphics[width=0.32\textwidth]{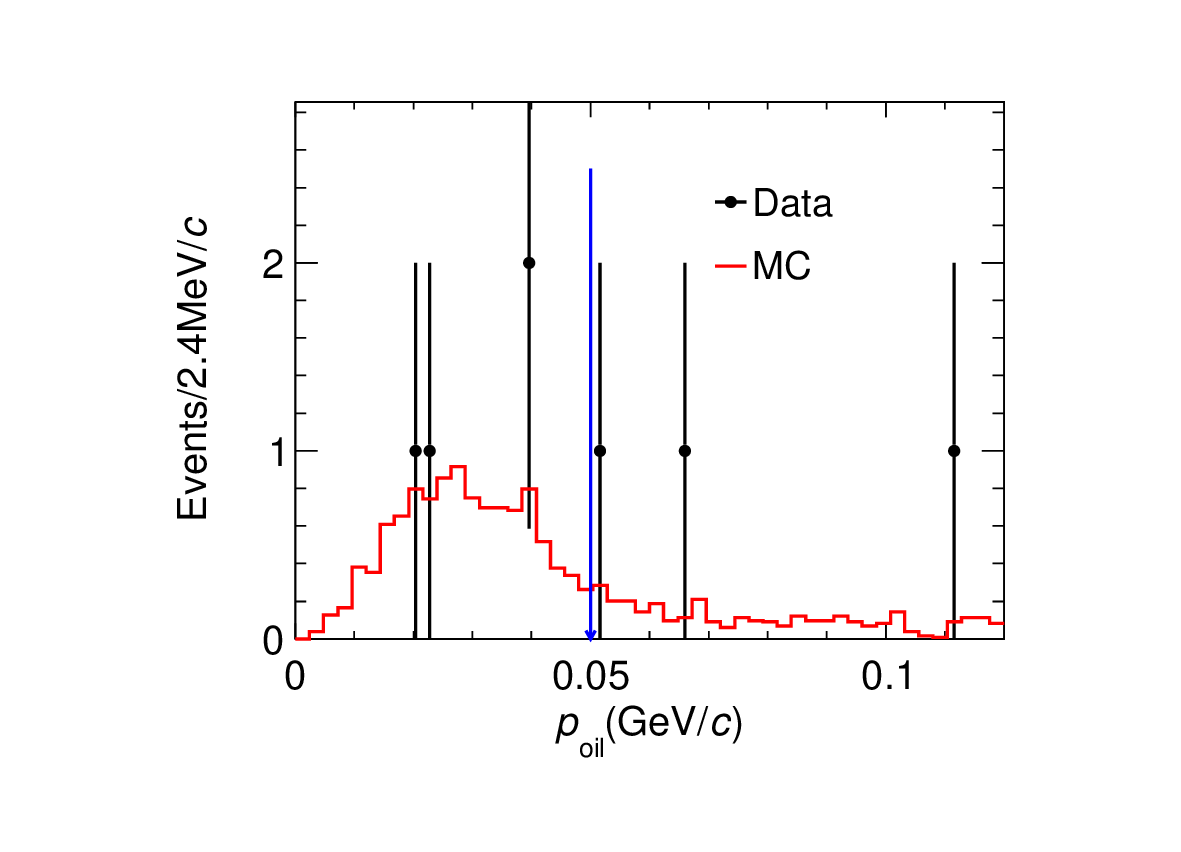}
    \end{center}
    \caption{Distributions of $p_{\mathrm{oil}}$ for the three studied $\bar{\Lambda}p \to K^+ \pi^+ \pi^- + k\pi^0$ channels (from left to right: $k=1,2,3$). Black points with statistical uncertainties represent data, and the red histograms correspond to MC simulations of quasi-at-rest $\bar{\Lambda}p$ annihilation on hydrogen protons in the oil layer, including the full detector response. The blue vertical line marks the applied requirement $p_{\mathrm{oil}} < 50~\mathrm{MeV}/c$.}
    \label{POil}
\end{figure*}

Signal processes $\bar{\Lambda} p \to K^+ \pi^+ \pi^- + k\,\pi^0$ are identified from the remaining charged tracks and $\pi^0$ candidates
after reconstructing the tagged $\Lambda$ candidates. 
Particle identification (PID) for these charged tracks combines measurements of the specific ionization energy loss in the MDC ($\mathrm{d}E/\mathrm{d}x$)
and the flight time in the time-of-flight system to form likelihoods $\mathcal{L}(h)(h = p, K, \pi)$ for various hadron hypotheses $h$.
Charged tracks with $\mathcal{L}(K)>\mathcal{L}(\pi)$ are identified as kaons, and pions are required to satisfy $\mathcal{L}(\pi)>\mathcal{L}(K)$.
A vertex fit is performed on all the $K^+ \pi^+ \pi^-$ combinations to obtain the vertex position of the $\bar{\Lambda} p$ interaction.
If there are multiple $K^+ \pi^+ \pi^-$ candidates, only the one with the smallest $\chi^2$ is retained for further analysis.
To select the best $\pi^0$ candidate for the three signal topologies,
we minimize $\chi^2_{1\mathrm{C}}(\pi^0)$ for $k=1$,
$\chi^2_{1\mathrm{C}}(\pi^0_1)+\chi^2_{1\mathrm{C}}(\pi^0_2)$ for $k=2$,
and $\chi^2_{1\mathrm{C}}(\pi^0_1)+\chi^2_{1\mathrm{C}}(\pi^0_2)+\chi^2_{1\mathrm{C}}(\pi^0_3)$ for $k=3$.
In case of multiple candidates, they are labeled as $\pi^0_i$ ($i=1,2,3$).
To select candidate events with an annihilation vertex in the beam pipe region 
and suppress background from non-$\bar{\Lambda}p$ interactions, the transverse distance of the 
reconstructed $K^+\pi^+\pi^-$ vertex from the beam axis ($R_{xy}$) is required to be within 
[2.9, 3.6]~cm.
%To select signal events that react with the cooling oil in the beam pipe, the distance of the reconstructed $K^+ \pi^+ \pi^-$ vertex position
%in the plane perpendicular to the beam direction($R_{xy}$) is required to be within  [2.9, 3.6]~cm.
The three-momentum of the struck proton is inferred from momentum conservation within the reconstructed system, as
\begin{equation}
\vec{p}_p = \vec{p}_{\mathrm{final}} - \vec{p}_{\bar{\Lambda}},
\end{equation}
with $\vec{p}_{\mathrm{final}} = \sum_i \vec{p}_i$ the vector sum of the reconstructed signal final-state momenta, the $\bar{\Lambda}$ momentum $\vec{p}_{\bar{\Lambda}}$ is taken as opposite to that of the reconstructed $\Lambda$. For a bound proton inside a nucleus, this inferred quantity corresponds to the Fermi momentum. For a free hydrogen proton in the oil layer, it is expected to vanish within experimental resolution.
The events from background reactions of the $\bar{\Lambda}$ antihyperon on the $^{197}\mathrm{Au}$/$^{9}\mathrm{Be}$/$^{12}\mathrm{C}$ nuclei
should have $p_{\text{oil}}$ (the magnitude of $\vec{p}_p$) in hundreds of MeV/$c$ due to the proton Fermi momentum in the nuclei,
in contrast to the very small $p_{\text{oil}}$ for the protons at rest in the signal processes.
By requiring $p_{\rm{oil}}<0.05$~GeV/$c$, annihilations with protons in these composite nuclei are minimized, and the $p_{\rm {oil}}$ distributions are shown in Fig~\ref{POil}.

\begin{figure}[htbp]
    \begin{center}
        \begin{overpic}[width = 1.0\linewidth]{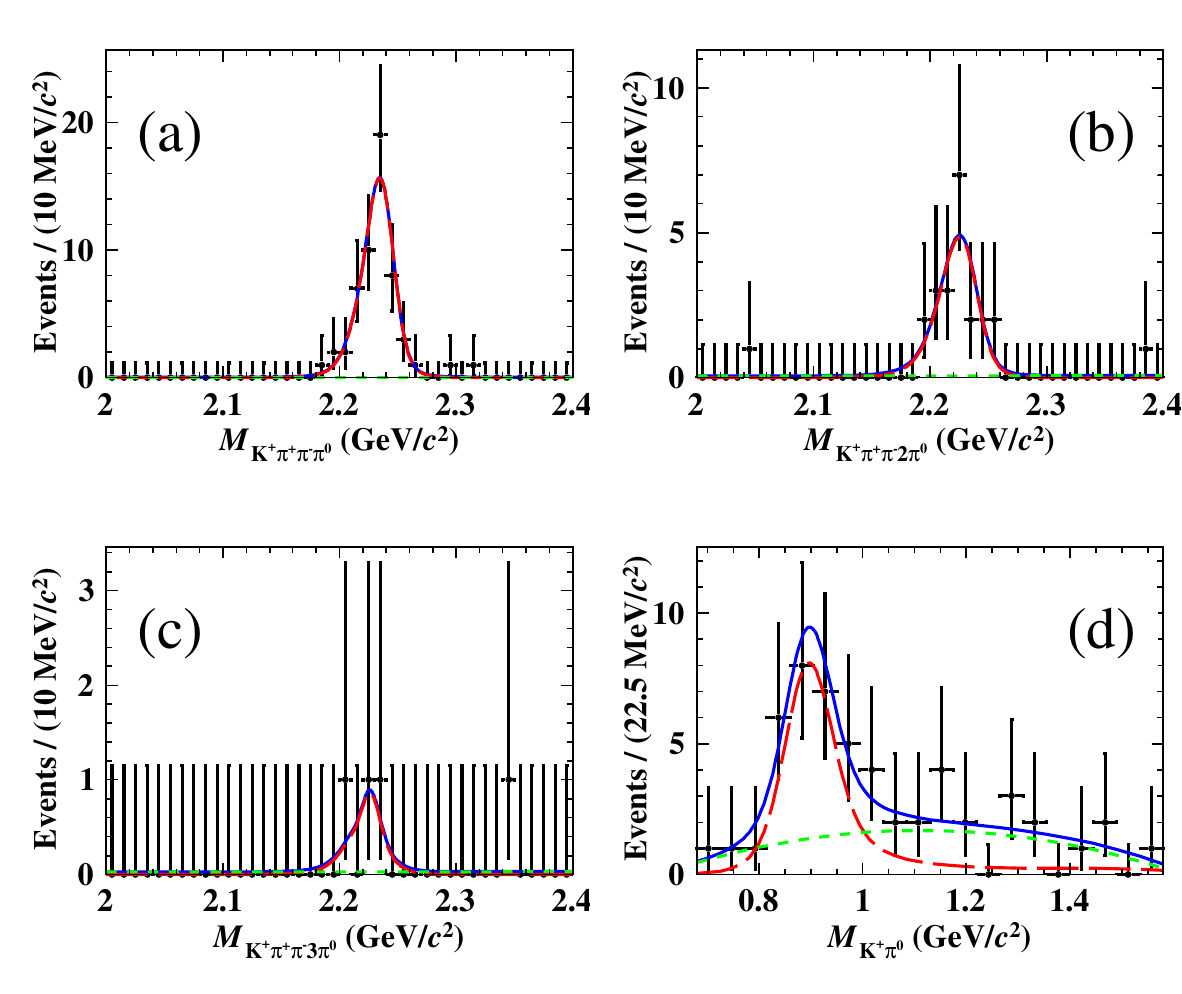}
        \end{overpic}
    \end{center}
    \vspace{-0.7cm}
    \caption{
        Fits to the $M_{K^+ \pi^+ \pi^- k\pi^0}$ (a)-(c) and $M_{K^+\pi^0}$ (d) distributions for the SA samples.
        The dots with error bars are data, the blue solid lines are the total fits,
        the red long-dashed and green short-dashed lines denote the fitted signal and background shapes, respectively.
    }
    \label{fig:SA}
\end{figure}

For signal events from $J/\psi\to\Lambda\bar{\Lambda}$,
the center-of-mass energy of an incident $\bar{\Lambda}$ annihilating on a stationary proton is effectively fixed
at $\sqrt{s}\simeq 2.243\pm0.005\ \mathrm{GeV}$, resulting in a peak near $2.243\ \mathrm{GeV}/c^2$ in the invariant mass spectrum of $K^+ \pi^+ \pi^- + k\,\pi^0$
($M_{K^+\pi^+\pi^-k\pi^0}$).
A detailed study of the inclusive $J/\psi$ MC sample shows no peaking background in the signal region.
To determine the signal yields, unbinned maximum-likelihood fits are performed to the $M_{K^+ \pi^+ \pi^- k\pi^0}$ distributions,
as shown in Figs.~\ref{fig:SA}(a)–(c).
In the fits, the signal is described with an MC-simulated shape convolved with a Gaussian function,
while the background is modeled with a first-order Chebyshev polynomial.
The SA yields are determined to be $N^{k=1}_{\rm SA}=53.2^{+7.8}_{-7.1}$, $N^{k=2}_{\rm SA}=20.5^{+4.9}_{-4.3}$,
and $N^{k=3}_{\rm SA}=2.9^{+2.1}_{-1.4}$, respectively, here the errors are statistical only.
The corresponding signal efficiencies,
estimated from the signal MC samples, are $\epsilon^{k=1}_{\rm sig}=20.77\%$, $\epsilon^{k=2}_{\rm sig}=8.71\%$, and $\epsilon^{k=3}_{\rm sig}=3.19\%$.
The signal significances for the $k=1$, $2$, and $3$ channels are $12.9\sigma$, $6.9\sigma$, and $2.1\sigma$, respectively,
evaluated from the change in the log-likelihood between fits with and without the signal component,
accounting for the associated change in the number of degrees of freedom.
The systematic uncertainties (which will be discussed later) are incorporated into the reported significance by adopting a conservative approach: the minimum significance observed across the systematic variations is quoted. 
%Systematic uncertainties are included in the significance calculations, which will be discussed later.

In the $k=1$ channel, evidence for the $K^{*}(892)^{+}$ resonance is observed in the $K^+\pi^0$ invariant mass spectrum $M_{K^+\pi^0}$,
as shown in Fig.~\ref{fig:SA}(d).
The yield of $\bar{\Lambda} p \to K^{*}(892)^{+}\,\pi^+\pi^-$ is determined to be $N^{K^{*}(892)^{+}}_{\rm SA} = 27.3^{+8.3}_{-7.4}$
from an unbinned maximum-likelihood fit,
in which the signal is described by the MC-simulated shape convolved with a Gaussian function,
and the background is modeled with a second-order polynomial.
The corresponding signal efficiency and significance are $\epsilon^{K^{*+}}_{\rm SA} = 21.77\%$ and $3.2\sigma$, respectively.
Due to limited statistics, the potential interference effects are neglected in the extraction of the signal yield.
Searches for other possible intermediate states (e.g. $K^*(892)^0$) are also performed, but no statistically significant signal is observed with the present data sample, preventing a quantitative discussion of possible isospin effects.

Based on Eq.~\ref{equ:crosssection},
the cross sections of the reactions $\bar{\Lambda} p \to K^+ \pi^+ \pi^- + k\,\pi^0 (k=1,2,3)$
are determined for the first time and found to be $\sigma_{\bar{\Lambda} p \to K^+ \pi^+ \pi^-\pi^0} = 8.5^{+1.2}_{-1.1}$~mb,
$\sigma_{\bar{\Lambda} p \to K^+ \pi^+ \pi^- 2\pi^0} = 7.9^{+1.9}_{-1.7}$~mb,
and $\sigma_{\bar{\Lambda} p \to K^+ \pi^+ \pi^- 3\pi^0} = 3.1^{+2.2}_{-1.5}$~mb
at the incident $\bar{\Lambda}$ momentum of $1.074 \pm 0.017~\mathrm{GeV}/c$, respectively.
For the subprocess $\bar{\Lambda} p \to  K^{*}(892)^{+}\,\pi^+\pi^-$,
Eq.~\ref{equ:crosssection} is modified by replacing the factor ${\mathcal B}^{k}_{\pi^0\to\gamma\gamma}$
with ${\mathcal B}_{\pi^0\to\gamma\gamma}\cdot{\mathcal B}_{ K^{*}(892)^{+}\to K^+\pi^0}$,
where ${\mathcal B}_{K^{*}(892)^{+}\to K^+\pi^0} = (33.300\pm0.003)\%$ is the branching fraction of the decay ${K^{*}(892)^{+}\to K^+\pi^0}$~~\cite{PDG},
yielding $\sigma_{\bar{\Lambda} p \to  K^{*}(892)^{+}\,\pi^+\pi^-} = 12.5^{+3.8}_{-3.4}\,\mathrm{mb}$.
Here, the uncertainties are statistical only.

Equation~\ref{equ:crosssection} shows that the uncertainties related to the ST efficiency cancel.
The sources of the systematic uncertainties are summarized in Table~\ref{tab:systematic}.
Each of them is described in the following.

\begin{table}[htbp]
  \caption{Relative systematic uncertainties (\%) of the measured
    cross sections.  $k=1$, $k=2$, $k=3$, and
    $K^{*}(892)^{+} \pi^+ \pi^-$ denote the signal processes
    $\bar{\Lambda} p \to K^+\pi^+\pi^- + k\pi^0$ and
    $\bar{\Lambda} p \to K^{*}(892)^{+}\pi^+\pi^-$, respectively. The
    symbol ``$\ldots$" indicates the systematic uncertainty from the
    corresponding source absent or negligible.}
    \label{tab:systematic}
    \setlength{\extrarowheight}{1.0ex}
     \renewcommand{\arraystretch}{1.0}
    \begin{center}
    \scalebox{0.91}{
        \begin{tabular} {l c c c c }
            \hline \hline
            Source & $~k=1~$ & $~k=2~$ &  $~k=3~$ & $~K^{*}(892)^{+} \pi^+ \pi^-$\\
            \hline
            Kaon tracking                   & $1.0$  & $1.0$ & $1.0$  & $1.0$      \\
            Kaon PID                        & $1.0$  & $1.0$ & $1.0$  & $1.0$      \\
            Pion tracking                   & $2.0$  & $2.0$ & $2.0$  & $2.0$      \\
            Pion PID                        & $2.0$  & $2.0$ & $2.0$  & $2.0$      \\
            $\pi^0$ reconstruction          & $1.0$  & $2.0$ & $3.0$  & $1.0$      \\
            ST signal shape                 & $1.3$  & $1.3$ & $1.3$  & $1.3$      \\
            ST background shape             & $1.1$  & $1.1$ & $1.1$  & $1.1$      \\
            SA signal shape                 & $0.6$  & $0.5$ & $\ldots$    & $3.7$      \\
            SA background shape             & $\ldots$    & $\ldots$   & $\ldots$    & $8.1$      \\
            $\alpha$ value                  & $0.2$  & $0.2$ & $0.2$  & $0.2$      \\
            Intermediate $K^{*}(892)^{+}$   & $2.1$  & $\ldots$   & $\ldots$    & $\ldots$        \\
            MC statistics                   & $1.0$  & $1.6$ & $2.6$  & $1.1$      \\
            \hline \hline
            Total                           & $4.4$  & $4.4$ & $5.4$ & $9.7$      \\
            \hline \hline
        \end{tabular}
    }
    \end{center}
\end{table}

The uncertainties arising from the tracking and PID are both 1.0\% per pion or kaon track~\cite{BESIII:2018ldc, BESIII:2024nif},
and the uncertainty due to $\pi^0$ reconstruction is 1.0\% per $\pi^0$~\cite{BESIII:2023ldd, BESIII:2023drj}.
The uncertainty associated with the ST signal shape of the $RM_{\Lambda}$ distribution is estimated by changing the model from
the Gaussian-convolved MC shape to the bare MC shape,
and the uncertainty from the ST background shape is evaluated by replacing the third-order Chebyshev polynomial with a second-order one.
The resulting changes in the measured cross sections are taken as the systematic uncertainties due to the ST signal and background shapes,
amounting to 1.3\% and 1.1\%, respectively, for all signal processes.
To investigate the uncertainties associated with the $R_{xy}$ and $p_{\rm oil}$ requirements,
the selection criteria are varied over a range of values, and the resulting impact on the cross sections is negligible.
To estimate the uncertainties from the SA signal shapes of the $M_{K^+\pi^+\pi^-k\pi^0}$ and $M_{K^+\pi^0}$ distributions,
the model is changed from the Gaussian-convolved MC shape to the bare MC shape.
The uncertainties associated with the SA background shapes for non-resonant channels($k=1,2,3$) are evaluated by replacing the first-order Chebyshev polynomial with a constant shape, which turn out to be negligible.
For the $\sigma_{\bar{\Lambda} p \to  K^{*}(892)^{+}\pi^+\pi^-}$ measurement, the background
shape is replaced by two alternatives: a third-order Chebyshev polynomial, and the $M_{K^+\pi^0}$ distribution from PHSP $\bar{\Lambda} p \to K^{+}\pi^+\pi^-\pi^0$ MC.
The larger of the resulting differences in the measured cross sections with respect to the nominal value is taken as the corresponding uncertainty.
The systematic effect of the interaction-point position in the MC simulation is found to be negligible by shifting
it by $\pm 0.2$~cm from the nominal position.
In the MC generator, the value of the angular-distribution parameter $\alpha$ from Ref.~\cite{BESIII:2022qax} is used.
To estimate its uncertainty, we vary $\alpha$ by $\pm 1\sigma$
and take the largest change in the detection efficiency as the systematic uncertainty.
In the nominal solution, the signal efficiency for the $k=1$ channel is determined using a PHSP $K^+\pi^+\pi^-\pi^0$ MC sample.
To assess the impact of the intermediate $K^{*}(892)^{+}$ component on the efficiency, we construct a mixed sample by combining
PHSP $K^+\pi^+\pi^-\pi^0$ MC with $K^{*}(892)^+\pi^+\pi^-$ MC. The ratio between these two MC samples is set according to their measured nominal cross sections when generating the mixed sample.
The difference between the nominal efficiency and that from the mixed sample, 2.1\%, is assigned as the systematic uncertainty.
The limited sample sizes of the MC samples are also considered a source of systematic uncertainty.
All systematic uncertainties are summarized in Table~\ref{tab:systematic},
and the total systematic uncertainty is obtained by adding the individual components in quadrature.

Table~\ref{tab:cross_section} summarizes the measured cross sections for the processes $\bar{\Lambda} p \to K^+\pi^+\pi^- + k\pi^0$ ($k=1,2,3$)
and $\bar{\Lambda} p \to  K^{*}(892)^{+}\pi^+\pi^-$, along with the corresponding signal significances.
Since the signal significance of the process $\bar{\Lambda} p \to K^+\pi^+\pi^-3\pi^0$ is less than $3\sigma$,
we also give the upper limit on its cross section at a 90\% confidence level (C.L.) using a Bayesian method, similar to the one employed in Ref.~\cite{BESIII:2024wvw}.
\begin{table}[htbp]
    \caption{The measured cross sections (or upper limit at 90\% C.L.) in mb.
    }
    \label{tab:cross_section}
    \setlength{\extrarowheight}{1.0ex}
     \renewcommand{\arraystretch}{1.0}
    \begin{center}
    \scalebox{1}{
        \begin{tabular} {l c c c c }
            \hline \hline
            Channel                                              & Cross section (upper limit)      \\
            \hline
            $\bar{\Lambda} p \to  K^+\pi^+\pi^-\pi^0$            & $8.5^{+1.2}_{-1.1}\pm0.4$                     \\
            $\bar{\Lambda} p \to  K^+\pi^+\pi^-2\pi^0$           & $7.9^{+1.9}_{-1.7}\pm0.4$                     \\
            $\bar{\Lambda} p \to  K^+\pi^+\pi^-3\pi^0$           & $3.1^{+2.2}_{-1.5}\pm0.2$ ($<7.2$)            \\
            $\bar{\Lambda} p \to  K^{*}(892)^{+}\pi^+\pi^-$      & $12.5^{+3.8}_{-3.4}\pm1.2$                    \\
            \hline \hline
        \end{tabular}
    }
    \end{center}
\end{table}

In summary, using $(10087 \pm 44)\times 10^6$ $J/\psi$ events collected with the BESIII detector,
we report the observation of the reactions $\bar{\Lambda}p \to K^+\pi^+\pi^- + k\pi^0$ with $k=1,2$,
and set an upper limit on the process $\bar{\Lambda}p \to K^+\pi^+\pi^- 3\pi^0$.
The $\bar{\Lambda}$ hyperons originate from $J/\psi \to \Lambda\bar{\Lambda}$ decays, traverse the beam pipe, and annihilate on the protons at rest in the cooling oil.
In addition, an evidence for an intermediate $K^{*}(892)^{+}$ in the $k=1$ channel is found in the invariant mass spectrum $M_{K^+\pi^0}$.
The cross sections of these signal processes are measured at $p_{\bar{\Lambda}}\simeq 1.074$~GeV/$c$ (corresponding to $\sqrt{s}\simeq2.243$~GeV),
as shown in Table~\ref{tab:cross_section}.
These measurements provide important quantitative constraints on the $\rm \bar{Y}N$ annihilation
into multi-meson final states, supply valuable input for transport and hadronization modeling in heavy-ion collisions,
including strange‑antibaryon survival and light‑antinuclei formation,
and enable tests of SU(3)-flavor symmetry via comparisons with existing $\bar{p} \rm N$ annihilation data.
Moreover, these data provide direct constraints on the imaginary part of the optical potential and 
are essential for testing the $G$-parity transformation in the strangeness sector.
The same approach can be pursued at the future Super $\tau-$Charm Facility~\cite{Achasov:2023gey} for more precise,
differential investigations of the $\rm \bar{Y}N$ scattering and annihilation dynamics.

% \clearpage
%% Saved at => 2025-10-14
\textbf{Acknowledgement}

The BESIII Collaboration thanks the staff of BEPCII (https://cstr.cn/31109.02.BEPC) and the IHEP computing center for their strong support. 
This work is supported in part by National Key R\&D Program of China under Contracts Nos. 2023YFA1606000, 2023YFA1606704; 
National Natural Science Foundation of China (NSFC) under Contracts Nos. 11635010, 11935015, 11935016, 11935018, 12025502, 12035009, 12035013, 12061131003, 12165022, 12192260, 12192261, 12192262, 12192263, 12192264, 12192265, 12221005, 12225509, 12235017, 12342502, 12342502, 12361141819; 
the China Postdoctoral Science Foundation under Grant No. 2024M753040;
the Postdoctoral Fellowship Program of China Postdoctoral Science Foundation under Grant No. GZC20241608;
the Chinese Academy of Sciences (CAS) Large-Scale Scientific Facility Program; the Strategic Priority Research Program of Chinese Academy of Sciences under Contract No. XDA0480600; 
CAS under Contract No. YSBR-101; 100 Talents Program of CAS; 
The Institute of Nuclear and Particle Physics (INPAC) and Shanghai Key Laboratory for Particle Physics and Cosmology; 
Yunnan Fundamental Research Project under Contract No. 202301AT070162;
ERC under Contract No. 758462; German Research Foundation DFG under Contract No. FOR5327; 
Istituto Nazionale di Fisica Nucleare, Italy; Knut and Alice Wallenberg Foundation under Contracts Nos. 2021.0174, 2021.0299, 2023.0315; 
Ministry of Development of Turkey under Contract No. DPT2006K-120470; 
National Research Foundation of Korea under Contract No. NRF-2022R1A2C1092335; 
National Science and Technology fund of Mongolia; 
Polish National Science Centre under Contract No. 2024/53/B/ST2/00975; 
STFC (United Kingdom); 
Swedish Research Council under Contract No. 2019.04595; 
U. S. Department of Energy under Contract No. DE-FG02-05ER41374.

%%%%%%%%%%%%%%%%%%%%%%%%%%%%%%%%%%%%%%%%%%%%%%%%%%%%%%%%%%%%%%%%%
%\bibliographystyle{apsrev4-1}
\bibliography{Reference.bib}
% \input{./LambarPro.bbl}

%%%%%%%%%%%%%%%%%%%%%%%%%%%%%%%%%%%%%%%%%%%%%%%%%%%%%%%%%%%%%%%%%
\end{document}